\documentclass[conference]{IEEEtran}
\IEEEoverridecommandlockouts
\usepackage{url}
\usepackage[noadjust]{cite}
\usepackage{amsmath,amssymb,amsfonts}
\usepackage{algpseudocode}
\usepackage{algorithm}
\usepackage{graphicx}
\usepackage{textcomp}
\usepackage{xcolor}
\usepackage{tikz}
\pdfoutput=1
\def\BibTeX{{\rm B\kern-.05em{\sc i\kern-.025em b}\kern-.08em
    T\kern-.1667em\lower.7ex\hbox{E}\kern-.125emX}}
\begin{document}

\title{Musical Chords: A Novel Java Algorithm and App Utility to Enumerate Chord-Progressions Adhering to Music Theory Guidelines\\

}

\author{\IEEEauthorblockN{Aditya Lakshminarasimhan}

}

\maketitle

\begin{abstract}
A song's backbone is its chord progressions, a series of chords that improve the harmony and add to the overall composition. For individuals ranging from beginners to creative artists, comprehending and implementing music theory grammar for their own compositions can stifle the music creation process and cause song-writer's block. The existing Chord Progression approaches in the marketplace are limited on producing only pre-selected progressions and often fail to conform to music theory guidelines or provide APIs for other musicians to build on. Because four-chord and eight-chord progressions are yet to be enumerated, Machine learning use-cases that train on chord progressions are limited, and mobile applications don't provide users with unique or unexplored progressions. To address these limitations, a novel Java Algorithm and automated music theory chord progression and variations generator App has been developed. This App offers a piano user interface, that applies music theory to generate all possible four-chord and eight-chord progressions and produces three alternate variations of the generated progressions selected by the user. The Algorithm elucidates 3,297 Total 4-Chord Progressions and 405,216 Total 8-Chord Progressions. Within the 4-Chord Progression pool, there are 1,533 Major 4-chord Progressions and 1,764 Minor 4-Chord Progressions. Within the 8-chord Progression pool, there are 182,094 Major Progressions and 223,122 Minor Progressions. This innovative approach provides musicians with a comprehensive and customizable tool for their music creation, allowing them to develop their signature sounds. 
\end{abstract}


\section{Introduction}
In a musical composition or work, a chord is a combination of 2 or more notes (single pitches). A foundational feature of classical and modern tonal music, a chord progression is a sequence of multiple chords. Chord progressions are fundamental to a song's structure as they guide the harmonic direction of a piece and the corresponding melody. As such, chord progressions in tonal music are highly prevalent across multiple music periods (Baroque, Classical, Romantic, and Contemporary) and genres such as classical music, jazz, rock, pop, and blues \cite{b1}.

Although chord progressions don't have an established sequence length, the most common lengths of chord progressions are 4-chord and 8-chord progressions \cite{b2}. Studies have shown that these sequences (relatively shorter than 16-chord or even 32-chord progressions) are more consonant and pleasing to listen to for the audience \cite{b3,b4}. This is attributed to the fact that after a progression concludes, it is common for the same progression to keep cycling throughout the remainder of a song. Often, the tension in the last chord of a progression will want to harmonically resolve back to the tonic (first chord in a tonic scale, denoted by I), starting the next sequence. The benefits of short sequences allow for more frequent and quicker scenarios of tense chords resolving to the tonic. This effect, in turn, is consonant and harmonically pleasing to the audience, allowing for effective music composition \cite{b4}. As such, 4-chord progressions and 8-chord progressions are among the most common progressions in the industry \cite{b5}.

Since their conception in the Baroque period of music (17th and 18th centuries), chord progressions have been used by elite musicians due to their ability to generate emotional and structural elements that define the trajectory of the piece. The process of generating chord progressions is a part of music theory, a grammar that defines the art of music with established guidelines. Extensive chord progression charts have been proposed over the fundamental periods of music composition, but the two most prevailing charts (major and minor charts) have guided the different progressions that can be created in each \cite{b6,b7}. In classical music, composers like Johann Sebastian Bach, Antonio Vivaldi, and Johann Pachelbel utilized complex chord progressions to create a sense of resolution and tension, which are essential in driving the narrative of their compositions \cite{b8,b9,b10,b11}. These progressions form the backbone of many classical pieces, providing a framework that supports melodic and rhythmic development. Similarly, contemporary musicians leverage chord progressions to craft memorable and emotionally resonant songs \cite{b12}. Throughout the history of music, these foundational chord progressions have been integrated by leading artists.

Research indicates a strong correlation between the use of effective chord progressions and commercial success in the music industry. Studies have shown that songs featuring well-structured chord progressions are more likely to achieve higher rankings on Billboard charts and other popular music success metrics. The familiarity and emotional resonance of these progressions play a significant role in their popularity among listeners \cite{b13,b14}.

Despite their importance, accessing and learning about chord progressions can be challenging for many musicians. Music theory knowledge, which allows for the proper conception and integration of chord progressions, bottlenecks the applications of generating unique chord progressions to the top musicians. Considering that in 2023, the total revenue of the recorded music industry amounted to 28.6 billion dollars and that there are estimated to be 35,520 industry-profession musicians, with countless more musicians who engage in music production as hobbies or for medical purposes like relaxation, effective access to all the chord progressions possible within the confines of four-chord and eight-chord progressions is crucial \cite{b15}. Numerous chord progression utilities \cite{b16,b17,b18} have been proposed to provide access to music progressions.

\section{Background Works}

With recent developments in Artificial Intelligence to generate chord progressions, impressive advancements have been made to improve user experience and design \cite{b19}. Two primary trends of research emerge in this space. The first primary trend in research conducted on music chord progressions involves Machine Learning approaches to generate chord progressions based on themes, user preferences, and other metrics. Reference \cite{b20}, for example, uses a variational autoencoder to generate chord sequences conditioned on the harmonic complexity of Western music. This approach considers metrics such as harmonic rhythm, harmonic dissonance, and harmonic evolution. Further approaches have treated chord progressions as a language model application \cite{b3,b21,b22}. As such, models that can classify chord progressions, generate chord progressions to fit a particular theme, and guide song creation have been proposed \cite{b23}.

These AI approaches are powerful in their ability to generate chord progressions that fit a particular style. However, there is yet to be an enumeration of every possible chord progression within the confines of music theory that can effectively help musicians choose unique chord progressions that haven't been explored. Additionally, treating chord progression generation as a classification issue, that fits chord progressions to a particular theme or time period of music, may only lead to chord progressions that fit within a common theme (i.e., using a 2-5-1 ending in a progression would classify the progression as a Jazz progression). Hence, the inherent classification of a generated chord progression to a specific genre is limiting and doesn't explore the full extent of chord progressions that are yet to be explored.

The second trend in research conducted on music chord progressions involves utilities that directly involve user interactions. One proposed approach, \cite{b16}, is particularly successful in its ability to document every standard chord in music theory (i.e., 1, 2, 3, ... n). Reference \cite{b16} also provides some chord progression sequences related to common Pop or Jazz progressions. Another approach, \cite{b17}, provides a utility that is capable of providing the user with chord sequences and rhythms for songs of their selection. Additional approaches \cite{b18} have proposed the ability to build chords and progression patterns through intuitive user-guided randomization. Furthermore, web applications \cite{b24} and \cite{b25} provide extensive knowledge on creating progressions and improving user connection. These developments have certainly improved user experience and design in the space of system design for musical applications. However, the approaches displayed above, along with many more on the iOS App Store, Google Play, and the Internet, inevitably suggest the same pool of progressions, which is a fraction of the permutations of progressions that are possible, so to speak. Specifically, these applications suggest chord progressions to a user that are within a particular genre (rock, pop, jazz, and classical). Due to the nature of the progression of music in the music industry, there is a very small pool of progressions in each genre due to established progressions showing success to a piece - which is also attributed to the fact that an enumeration of every possible chord progression, including unique progressions that haven't been explored, is yet to be conducted. Nonetheless, users are exposed to the same pre-selected pool of 10-15 chord progressions across various utilities, which have already been intricately studied by the most successful mainstream artists. Ultimately, it is difficult for musicians aspiring to break through in this competitive space due to the limitations of music theory knowledge and exposure to a small pool of progressions that have already been explored. Another key point of improvement in this space is that many of these approaches only provide the four-chord or eight-chord sequence for users to build their melodies. However, there is yet to be a tool that can dynamically provide suggestions for how to build on from that base chord progression. This is another feature of music theory knowledge that enables successful musicians to build onto their songs by enabling them to find which keys can use the same numerical progression and harmonically build off the tonic progression.

With consideration of the importance of chord progressions to the harmonic structure of a song, but also the existing limitations due to the lack of a utility that enumerates every possible four-chord and eight-chord progression, this paper proposes the following:

A novel Java Algorithm and approach to recursively enumerating every possible 4-chord and 8-chord Progression within music theory guidelines.
An intuitive mobile utility that can process the proposed batch progressions and map them onto a user interface.
An enumeration that reveals 157 total four-chord progressions and 405,216 total eight-chord progressions.
A feature on the mobile utility that suggests alternate scale progressions to the main progression based on scale-degree music theory and numerical progression.

\section{Methodology}
Chord Theory Charts, along with Chord Progressions themselves, were first conceptualized in the 17th century \cite{b26}. Chords were said to resolve their tensions through movements. Over hundreds of years, musicians have revised chord progressions on the basis of emotional response. from audiences. 

\begin{figure}[htbp]
\centerline{\includegraphics{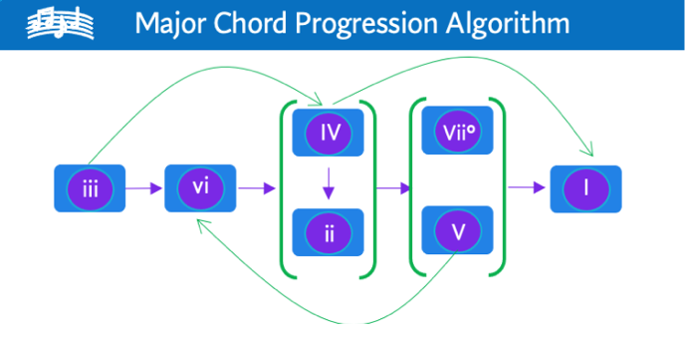}}
\centerline{\includegraphics{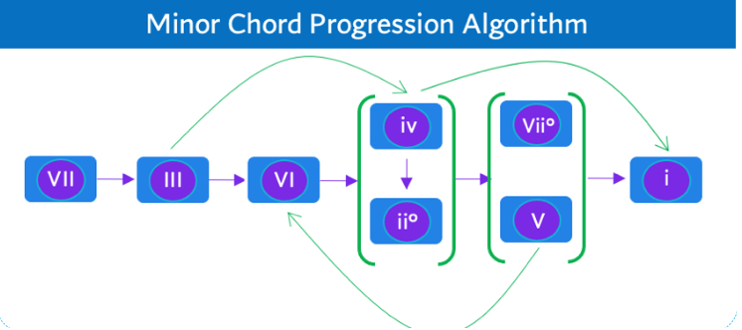}}
\caption{Major and Minor Chord Theory Charts. These charts were recreated using new visuals. These charts describe chord progression pattens and are applicable to both four-chord progressions and eight-chord progressions. The top diagram is applicable to the production of major-progression sequences and the bottom diagram is applicable to the production of minor-progression sequences. Each of the above Roman Numerals represents a numerical scale degree with a corresponding chord that can be applied depending on the user-selected scale. The constructed charts refer to \cite{b26} but also take design inspiration from \cite{b27}.}
\label{fig}
\end{figure}

The process of creating a chord progression is notably illustrated in Chord Theory Charts. Chord Theory charts are a map with chords and arrows that describe the progression or ‘movement’ from one chord to another. As there are two base-classifications in music – Major scales and Minor Scales – there are two Chord Theory charts: one distinct chart for each group. 

\subsection{Major, Minor, and Diminished Chords}
The two broad classifications of chords fall into Major chords and Minor Chords. In music theory, a major chord is a combination of a root (tonic - first scale degree), major third (relative to tonic - first scale degree), and perfect fifth (relative to tonic - first scale degree). A major third interval relative to the tonic is 4 half steps above the tonic in a standard 12 half step scale. A perfect fifth interval relative to the tonic is 7 half steps above the tonic in a standard 12 half step scale. Another method of rapidly identifying major chords, is that major chords are made up of one Major Third interval and one Minor Third interval. The interval between the root and the second note should be a Major Third, while the interval between the second note and the third note should be a Minor Third. 

Similarly, minor chords are chords that stem from the root (tonic of a scale). The minor chord is a combination of a root (tonic - first scale degree), minor third (relative to tonic - first scale degree), and perfect fifth (relative to tonic - first scale degree). A minor third interval relative to the tonic is 3 half steps above the tonic in a standard 12 half step scale. A perfect fifth interval, similarly, is 7 half steps above the tonic in a standard 12 half step chromatic scale. Similar to major chords, there is a method to rapidly identify minor chords: minor chords are made up of one Minor Third interval and one Major Third interval. The interval between the root and the second note should be a Minor Third, while the interval between the second note and the third note should be a Major Third. 

Diminished chords represent the third type of chords denoted on the chart. These chords are a combination of a root (tonic - first scale degree), minor third (relative to tonic - first scale degree), and diminished fifth (relative to tonic - first scale degree). The process to building a diminished chord is as follows: The interval between the root and the second note should be a Minor Third, while the interval between the second note and the third note should be another Minor Third. 

\subsection{Scale Degrees and Notes}

In music theory, a scale is any set of musical notes ordered by fundamental frequency or pitch. Musical scales in traditional Western music consist of seven notes and repeat at the octave. As such, each note (named 'Scale Degree') in a scale is assigned an index from 1 through 7. In Fig. 1, the top chart representing the 'Major Chord Progression Algorithm' consists of numbers 1 through 7 where each numerical scale degree represents a three note chord that is built off of that scale degree. For example, employing the major chord progression algorithm in the scale of C-Major would consist of scale degrees C (I), D (ii), E (iii), F (IV), G (V), A (vi), B (vii°). The 'I' on the major chart would represent a three-note chord built off C. This chord would be C, E, G. 

In Fig. 1, the second chart titled 'Minor Chord Progression Algorithm' consists of a slight alteration to the major chart: scale degrees are indexed from 1 through 7, however, there are two variants of the 7 chord. Specifically, the two types of 7 numerical chords are 'VII' and 'vii°'. In Western Music, the two variants of the 7th chord in a scale represent major 7th chords and diminished 7th chords respectively. Major 7th chords follow the typical major chord pattern, in which there is a major third interval and a perfect fifth interval built off of the 7th note in the scale. Diminished 7th chords, however, are composed of a minor third interval and a diminished fifth interval. 

\subsection{Reading and Interpreting the Chord Theory Charts}
Uppercase Roman Numerals (I, II, III, IV, V, VI, VII) represent Major chords while Lowercase Roman Numerals (i, ii, iii, iv, v, vi, vii) represent Minor chords. Roman Numerals with a '°' denote diminished chords (i.e. ii° and vii°). Both charts in Fig. 1 can be read to produce chord progressions. In music theory, the first chord used in a chord progression is the tonic, or the 'I' chord. Generally, the next chord in a sequence is where the arrow from the current chord points to. 

For major progressions the first chord must be a 'I' and for minor progressions the first chord must be a 'i'. From the 'I' and 'i' chord (tonic of major and minor respectively) any of the other chords on the respective chord progression algorithm charts in Fig. 1 are allowed as the next chord in the sequence (the I and i chord can be followed by any chord 1 through 7). Generally, the process to creating a chord progression is as follows: Start with the Tonic Chord; Choose the second chord in the progression (1 through 7). Depending on the second chord of the progression, the third chord will be where the arrow from the second chord points to. Depending on the third chord of the progression, the fourth chord will be where the arrow from the third chord points to. This process carries through for the remainder of the chords depending on the sequence length of a progression (i.e. 4-chord and 8-chord). This process is recursive in nature and is described by a set of rules, that is the next 'allowed' chord in a sequence. As such, these rules can be recreated in a coding language to enumerate every progression. 

In both cases, arrows are not bidirectional. A next chord in a sequence can only be chosen based on the direction that the arrow points. In both cases, a chord in the first set of brackets is able to go to either chord in the second set of brackets. The reverse, however, is not necessarily permitted. In a given bracket, a chord can only go to another chord in that same bracket if an arrow points in the direction of that chord (i.e. IV to ii is permitted, but vii° to V is not). 

For Major Chord Progressions:
\begin{itemize}
\item The 'I' chord is allowed to go to any of the chords 'I' through 'vii°' 
\item The 'ii' chord is allowed to go to either 'vii°' or 'V'
\item The 'iii' chord is allowed to go to either 'IV' or 'vi'
\item The 'IV' chord is allowed to go to either 'ii', 'vii°', 'V', or 'I' 
\item The 'V" chord is allowed to go to either 'vi' or 'I' 
\item The 'vi' chord is allowed to go to either 'IV' or 'ii' 
\item The 'vii°' can only go to 'I"
\end{itemize}

For Minor Chord Progressions:
\begin{itemize}
\item The 'i' chord is allowed to go to any of the chords 'i' through 'vii°/VII' 
\item The 'ii°' chord is allowed to go to either 'vii°' or 'V'
\item The 'III' chord is allowed to go to either 'iv' or 'VI'
\item The 'iv' chord is allowed to go to either 'ii°', 'vii°', 'V', or 'i'
\item The 'V" chord is allowed to go to either 'VI' or 'i' 
\item The 'VI' chord is allowed to go to either 'iv' or 'ii°' 
\item The 'vii°' can only go to 'i"
\item The 'VII' can go to 'III'
\end{itemize}

\subsection{Ionian Major and Minor Scales}
In Western Music, there are 7 modes that make up fundamental music composition: Ionian, Dorian, Phrygian, Lydian, Mixolydian, Aeolian, and Locrian. Each major mode is an alteration of the default/standard Ionian Mode, while each Minor mode is an alteration of the default Aeolian Mode. For example, the Dorian Mode raises the 6th scale degree of the Ionian Scale by a half step, warranting an entirely different tonal texture and mode. On the other hand, the Phrygian mode lowers the 2nd scale degree of the Aeolian Mode by a half step. Numerical progressions remain constant across all modes, as such, altering a mode will not require a different numerical progression or change the process of creating one as described in Fig. 1. 

Most modern music, including a variety of musical styles, utilize the Ionian Mode of music for Major Scales and the Aeolian Mode of music for Minor Scales. 

\subsection{Combining Scale with Numerical Progression}
Given that the Major and Minor Chord Progression Algorithm Charts in Fig. 1 generate numerical progressions, these progressions are generic to any scale. To make these scales applicable to songs and to make them easier to understand for beginner musicians, implementing combinations of scales and modes is important. Each note in the default scale (C, D, E, F, G, A, B) can have 3 different notes.
In the example of C, there can be natural 'C', 'C Sharp', and 'C Flat'. This indicates that each of the 7 notes in a scale have these 3 types leading to 21 different scales. Granted, there are 12 half steps in a scale, there is a degree of overlap. For example, C Flat and B Natural refer to the same pitch, rendering them as 'enharmonic'. While these overlap pitches may sound the same, they are notated as different pitches. These differences depend on what scale the user has selected, which can inevitably have impacts for other instruments where the scale at hand translates to a part of playing those instruments. In these cases, enharmonic pitches are non-negligible and should be represented in the full 21 scale format, regardless of overlapping pitches.

In Ionian Scales (typically Major Scales) the general formula can be written as (where Whole steps 'W' represent 2 half steps): 
\[
\begin{array}{c|c|c|c|c|c|c|c}
\text{Note} & 1 & 2 & 3 & 4 & 5 & 6 & 7 \\
\hline
\text{Interval} & W & W & H & W & W & W & H \\
\end{array}
\]
In Aeolian Scales (typically Minor Scales) the general formula can be written as: 
\[
\begin{array}{c|c|c|c|c|c|c|c}
\text{Note} & 1 & 2 & 3 & 4 & 5 & 6 & 7 \\
\hline
\text{Interval} & W & H & W & W & H & W & W \\
\end{array}
\]

\section{Proposed Java Algorithm}
The movement of chords from one to another as described above, can be denoted as a set of 'rules'
 that can be programmed. 

\subsection{Equation and Logic}
\begin{itemize}
\item \( L \) is the length of the chord progression.
\item \( C_i \) represents the chord at position \( i \).
\end{itemize}

The set of possible transitions \(T_{\text{Major}}(C_i) \) for Major Chords can be defined as:
\[
T_{\text{Major}}(C_i) =
\begin{cases}
\{ii, iii, IV, V, vi, vii^\circ\}, & \text{if } C_i = I \\
\{V, vii^\circ\}, & \text{if } C_i = ii \\
\{IV, vi\}, & \text{if } C_i = iii \\
\{ii, V, I, vii^\circ\}, & \text{if } C_i = IV \\
\{vi, I\}, & \text{if } C_i = V \\
\{ii, IV\}, & \text{if } C_i = vi \\
\{I\}, & \text{if } C_i = Vii^\circ \\
\end{cases}
\]
The set of possible transitions \(T_{\text{Minor}}(C_i) \) for Minor Chords can be defined as:

\[
T_{\text{Minor}}(C_i) =
\begin{cases}
\{ii^\circ, III, iv, V, VI, vii^\circ, VII\}, & \text{if } C_i = i \\
\{V, vii\circ\}, & \text{if } C_i = ii^\circ \\
\{iv, VI\}, & \text{if } C_i = III \\
\{ii^\circ, V, i, vii^\circ\}, & \text{if } C_i = iv \\
\{VI, i\}, & \text{if } C_i = V \\
\{ii^\circ, iv\}, & \text{if } C_i = VI \\
\{i\}, & \text{if } C_i = vii^\circ \\
\{III\}, & \text{if } C_i = VII \\
\end{cases}
\]

Starting with \( C_1 = I \) (For major chords 'I' and for minor chords 'i', the chord progression equation for \( L \) steps can be written as:
\[
\begin{aligned}
C_1 &= I \quad \text{or} \quad i \\
C_2 &\in T_{\text{Major}}(C_1) \quad \text{or} \quad T_{\text{Minor}}(C_1) \\
C_3 &\in T_{\text{Major}}(C_2) \quad \text{or} \quad T_{\text{Minor}}(C_2) \\
C_4 &\in T_{\text{Major}}(C_3) \quad \text{or} \quad T_{\text{Minor}}(C_3) \\
&\vdots \\
C_L &\in T_{\text{Major}}(C_{L-1}) \quad \text{or} \quad T_{\text{Minor}}(C_{L-1})
\end{aligned}
\]

\begin{equation}
C_i = 
\begin{cases}
I \,/\, i, & \text{if } i = 1 \\
C_{i+1} \in T_{\text{Major}}(C_i) \,/\, T_{\text{Minor}}(C_i), & \text{for } 2 \leq i \leq L
\end{cases}
\label{eq:chord-progression}
\end{equation}

Equation (1) is a representation of how the Java Algorithm is built at a High Level. Here, \( C_i \) is defined iteratively starting from \( I \), with each subsequent chord \( C_{i+1} \) selected from the set \( T_{\text{Major}}(C_{i}) \quad \text{or} \quad T_{\text{Minor}}(C_{i})\). The starting chord of the progression (\({i} \) = 1) can be either the I or i chord. For the purpose of enumeration, the first iteration of the algorithm goes through with setting  \( C_i \) = \(I\) given \({i} \) = 1 for Major progressions and generates all progressions within that scope. Similarly, in the next iteration of the enumeration, the algorithm goes through with setting  \( C_i \) = \(i\) given  \({i} \) = 1 for Minor Progressions and generates all progressions within that scope. The next chord in the progression \( C_{i+1} \), when the Major selection is made, will generate the remaining \(L\) chords to complete the major progression. Similarly the next chord in the progression for \( C_{i+1} \) when the Minor selection is made, will generate the remaining \(L\) chords to complete the minor progression. 

\subsection{Java Algorithms for Numerical Progressions}
\begin{algorithm}
\caption{Major Chord Progression Java Algorithm}
\begin{algorithmic}[1]
\If {Sequence == 'M'}
    \State Define Transitions as: $T_{\text{Major}}(C_i)$
    \State Initialize MajorCounter to 0
    \For{$N_2$ from 1 to 7}
        \State Define $N_3$ as \( T_{\text{Major}}(N_2) \)
        \For{c from 1 to length of $N_3$}
            \State Define $N_4$ as \( T_{\text{Major}}(N_3[c]) \)
            \For{d from 1 to length of $N_4$}
                \State Define $N_L$ as \( T_{\text{Major}}(N_{L-1}[d]) \)
                \State Print ``1, $N_2$, $N_3[c]$, $N_4[d]$, $N_L[d]$, Major''
                \State Increment MajorCounter
            \EndFor
        \EndFor
    \EndFor
    \State Print ``Total Possibilities: '', MajorCounter
\EndIf
\end{algorithmic}
\end{algorithm}
Algorithm (1) represents the Java logic that enumerates all numerical Major four-chord and eight-chord progressions where L = 4 and L = 8 respectively. Each nested for-loop section considers the selected chord of the previous for-loop to ensure that the appropriate pool of chords from \(T_{\text{Major}}(C_i) \) is chosen (i.e. If the 2nd For-Loop selects the \(ii\) chord of the scale, only the \(V\) or \(vii^\circ\) chords are accessible for the 3rd For-Loop). Fig. 2 refers to the combination of numerical progressions from the algorithm with scales.

\begin{algorithm}
\caption{Minor Chord Progression Java Algorithm}
\begin{algorithmic}[1]
\If {Sequence == 'm'}
    \State Define Transitions as: $T_{\text{Minor}}(C_i)$
    \State Initialize MinorCounter to 0
    \For{$N_2$ from 1 to 8}
        \State Define $N_3$ as \( T_{\text{Minor}}(N_2) \)
        \For{c from 1 to length of $N_3$}
            \State Define $N_4$ as \( T_{\text{Minor}}(N_3[c]) \)
            \For{d from 1 to length of $N_4$}
                \State Define $N_L$ as \( T_{\text{Minor}}(N_{L-1}[d]) \)
                \If {$N_2$ == 8}
                    \State Print ``1, 7Maj, $N_3[c]$, $N_4[d]$, Minor''
                \ElsIf {$N_3[c]$ == 8}
                    \State Print ``1, $N_2$, 7Maj, $N_4[d]$, Minor''
                \ElsIf {$N_4[d]$ == 8}
                    \State Print ``1, $N_2$, $N_3[c]$, 7Maj, Minor''
                \Else
                    \State Print ``1, $N_2$, $N_3[c]$, $N_4[d]$, Minor''
                \EndIf
                \State Increment MinorCounter
            \EndFor
        \EndFor
    \EndFor
    \State Print ``Total Possibilities: '', MinorCounter
\EndIf
\end{algorithmic}
\end{algorithm}
\begin{figure}[htbp]
\centerline{\includegraphics[width=0.8\columnwidth]{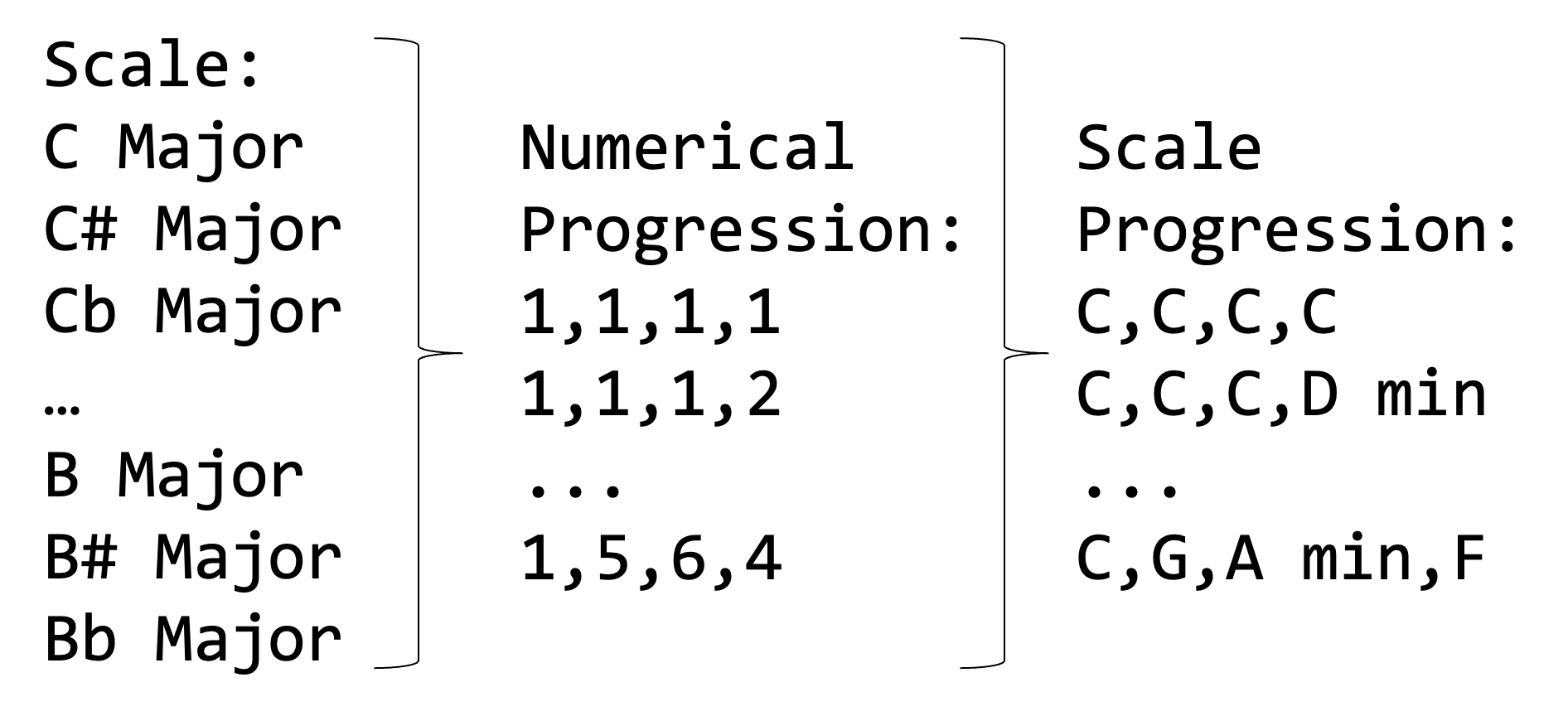}}
\caption{Process Diagram of Java Algorithm translating Numerical Progressions into Scale Progressions that are display-ready for users. Examples of C...Bb Major are provided, but this method is used for all 21 Major and Minor Scales.}
\label{fig}
\end{figure}

An almost identical algorithm is proposed for the Minor Chord Progression Java Algorithm with one key distinction. The Minor Chord Progression Algorithm denotes the \(VII\) chord with the number '8' and the \(vii^\circ\) chord with the number '7' for ease of distinction and interpretability. Given that the natural minor scale and the default major scale both denote the \(vii^\circ\) chord with the roman numeral for '7', a similar process is outlined with the Minor Progression Algorithm. 

Upon utilizing the Java Algorithms, the four-chord progressions and eight-chord progressions were successfully enumerated. This output was first generated in the console and then automatically transcribed to a csv using the Java PrintWriter class. Given that this Java Algorithm was successfully able to auto-generate every numerical four-chord and eight-chord progression within Music Theory Grammatical bounds, the next step was to translate the numerical chord sequence into a usable chord sequence based on Ionian scales.
\subsection{Conversion of Numerical Progression to Scale Progression}
Upon successful completion of enumerating all possible (within the confines of the Music Theory Charts in Fig. 1) numerical chord progression, the next step involves combining each progression with each of the 21 Ionian Major scales and 21 Aeolian Minor Scales. As illustrated in Fig. 2, the process involves taking 1 scale out of the pool of 21 Major Progressions and running each of the numerical Progressions for that scale to generate the scale progression. Given that each number in a numerical chord progression is an index that corresponds to a specific chord in out of all chords in a scale, the relative numbers correspond to specific pitches on scales. Table I illustrates how numerical progressions are added with scales.

\section{Results}
\begin{table}[htbp]
\caption{Java Batch Algorithm Dataset}
\begin{center}
\begin{tabular}{|c|c|c|c|}
\hline
\textbf{Scale} & \textbf{Number Progression} & \textbf{Scale Progression}  \\
\hline
C-major & 1,1,1,1 & C,C,C,C \\
\hline
C\#-major & 1,1,1,1 & C\#,C\#,C\#,C\# \\
\hline
... & ... & ... \\
\hline
A-minor & 1,5,6,4 & Am,Em,F,Dm  \\
\hline
\end{tabular}
\label{tab1}
\end{center}
\end{table}

\subsection{Java Batch Algorithm Results}
The Java Algorithm was able to successfully enumerate every possible 4-chord and 8-chord progression with regards to the Theory method of producing Chord Progressions. The Algorithm elucidates 3,297 Total 4-Chord Progressions and 405,216 Total 8-Chord Progressions. Within the  4-Chord Progression pool, there are 1,533 Major 4-chord Progressions and 1,764 Minor 4-Chord Progressions. Within the 8-chord Progression pool, there are 182,094 Major Progressions and 223,122 Minor Progressions. This enumeration study sheds light on every possible chord progression within the confines of classical music theory guidelines. As such, the progressions include hundreds, if not, thousands of understudied and underutilized chord progressions. Existing AI capabilities that attempt to derive mood or feeling from progressions have been traditionally limited by a lack of chord progression data (pre-selected pool of progressions, some of which don't conform to music theory guidelines). With the enumeration of all of these progressions, Machine learning approaches can now train on a much larger set of data to explore connections or feelings that haven't yet been explored. 
\subsection{Integration of Alternate Variations}
Another key contribution of this paper, with a user-focussed engineering goal, is to provide customizability, ease of integration, and structure for song-writing to prevent song-writers block. To address the limitation of current strategies that provide chord progressions without any additional measures to help aspiring musicians build on to compositions, a novel alternate variation technique has been designed. In music theory, there are 2 prescribed methods of generating alternate variations for chord progressions - one for major and the other for minor. For Major Chord progressions, the alternate variations based on a selected scale can be designed by applying the numerical chord progression to the 4th, 5th, and 6th scale degrees of the starting scale.

For example, the scale of C Major is 'C, D min, E min, F, G, A min, B dim'. The 'main progression' for the numerical progression '1,1,1,1' would be 'C, C, C, C'. Alternate chord progressions would be built based on F(4) Major, G(4) Major, and A(4) Minor because these are the 4th, 5th, and 6th scale degrees of the C-Major scale and have been combined with the same numerical progression, '1,1,1,1'. 

For Minor chord progressions, the alternate variations based on a selected scale can be designed by applying the numerical progression to the 3rd, 4th, and 5th scale degrees of the starting minor scale. For example, the scale of A Minor is 'A min, B dim, C, D min, E min, F maj, G'. Alternate chord progressions would be based on C(3) Major, D(4) Minor, and E(5) Minor. 

\subsection{Mobile System Design}
Fig. 3 is a process diagram for the iOS Mobile System in which the chord progression data generated from the algorithm and alternate variations above, is packaged onto a User Interface. 
\begin{figure}[htbp]
\centerline{\includegraphics[width=1.0\columnwidth]{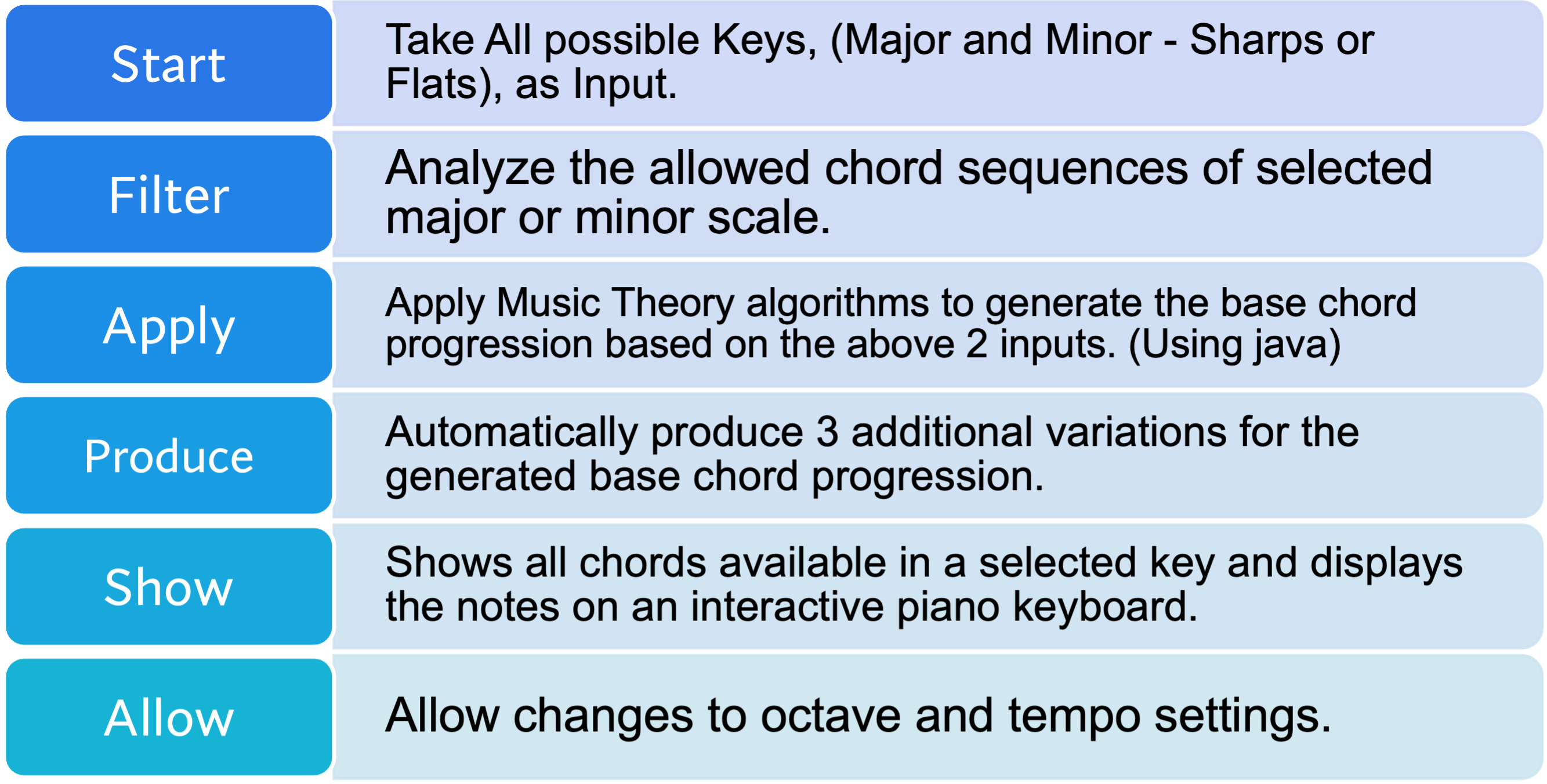}}
\caption{Process diagram for iOS Mobile App Design. Enumerated chord progressions from the Java Utility are used as inputs for the Mobile Application. }
\label{fig}
\end{figure}
To begin, the iOS Mobile System Application was developed in XCode using the interface language SwiftUI. Standard packages were used, no additional packages were imported. The version of XCode used is Xcode 11. Within the SwiftUI language, the Storyboard Interface language was used for real-time connectivity between the backend code and frontend display. The AVAudioEngine API provided from Apple was used to control sound, MIDI input, and tempo (which are all modified later in the design. 

To describe the process in Fig. 3, the iOS Mobile App design consists of 6 broad steps. 
The first step involves the starting display and inputs which takes all possible keys, (Major, Minor, Sharps, or Flats) as inputs. Next, the 'allowed chord sequences' of the user-selected major or minor chord are filtered. Specifically, a chord is only deemed as an 'allowed chord sequence' if it is part of the selected scale. The next step is applying music theory algorithms to generate the base chord progression based on the above inputs using the Java-enumerated progressions stored in the dataset. Next, the 3 additional variations from the Java Algorithm Dataset are produced for the selected base chord progression. Then, all chords available in a selected scale are displayed on the bottom panel of the User Interface, and an interactive piano automatically plays the chord progression. Finally, octave and tempo steppers are displayed for the user to modify and customize.

Fig. 4 is a screenshot of the iOS Mobile app designed in XCode using the SwiftUI Storyboard language. SwiftUI is an interface development language developed by Apple that connects the code to the display in real time. In order to successfully design a iOS Mobile App through XCode, elements such as tables, buttons, updating text fields, and data inputs must be dynamically mapped to the display such that the app can perform without latency. For design purposes, Vertical and Horizontal stacks along with constraints were used to position the visual elements. 

Inside Fig. 4, there are 4 different screenshots of the mobile app design. In each screenshot, a different functionality of the app is revealed. Specifically, on the opening of the app, users can select a tempo from a stepper ranging from 20 beats per minute to 300 beats per minute. Users can also select an octave stepper ranging from C1 to C9 scales respectively. After these scales and beats per minute, sound texture is distorted and of no use to the user. The scale and Progression text fields in the top panel are dynamic and display different chord progressions in a scroll wheel that appears from the bottom depending on major or minor scale selection. Specifically, the app currently works for 4 chord progressions and will display 73 numerical progressions upon 'Major selection' and 84 numerical progressions upon ''Minor selection'. 

The entire system is 'in-house' and uses the dataset from the Java utility to derive data. The data is then distributed to different sections of the app. Fig. 5 is a detailed description of the call hierarchy that powers the app. 
\begin{figure}[htbp]
\centerline{\includegraphics[width=1.0\columnwidth]{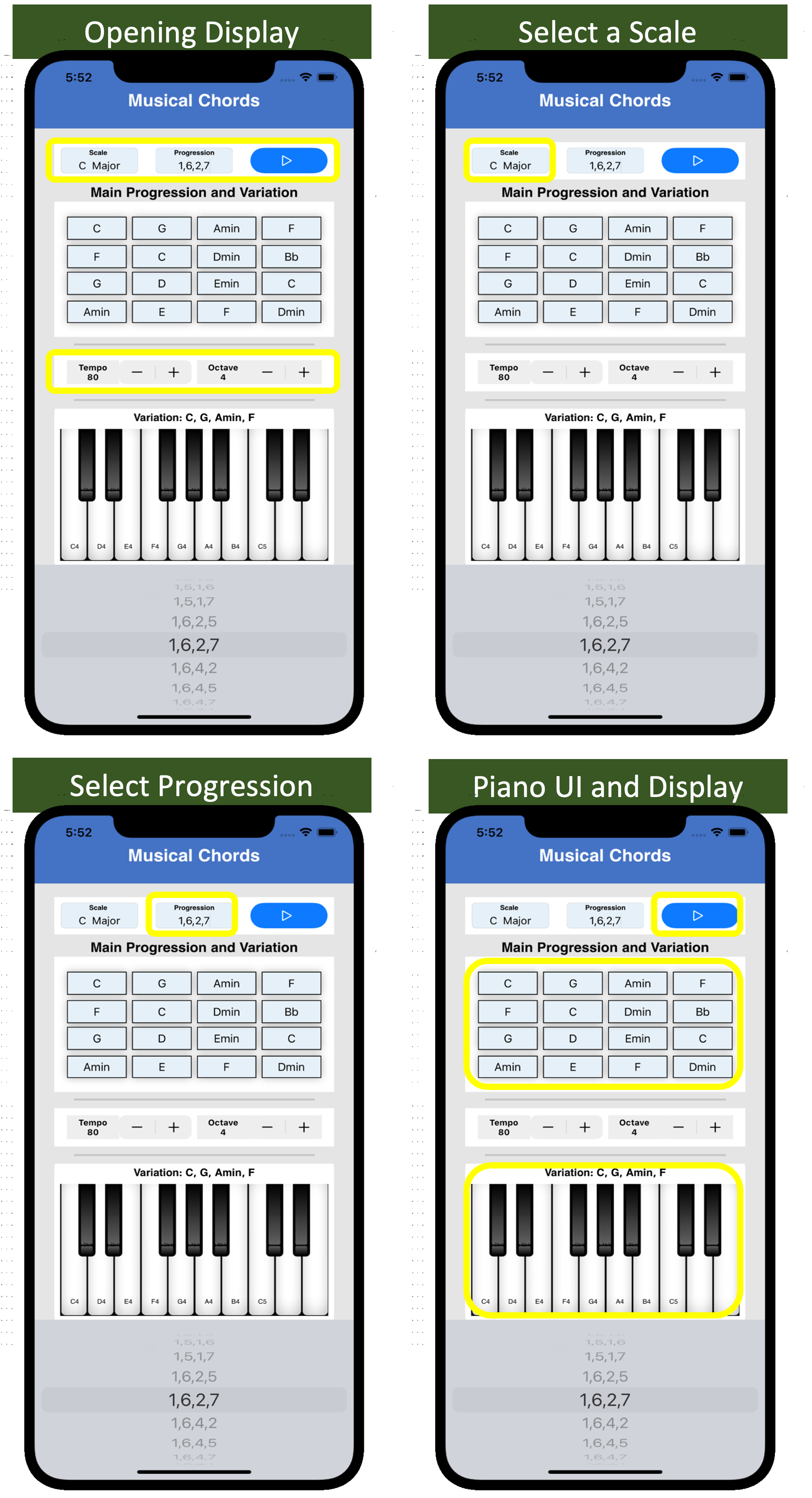}}
\caption{Four displays of the iOS Mobile app that show functionality. The screenshots were each taken with an iPhone 11 Simulation from XCode.}
\label{fig}
\end{figure}

\begin{figure*}[htbp]
\centerline{\includegraphics[width=\textwidth]{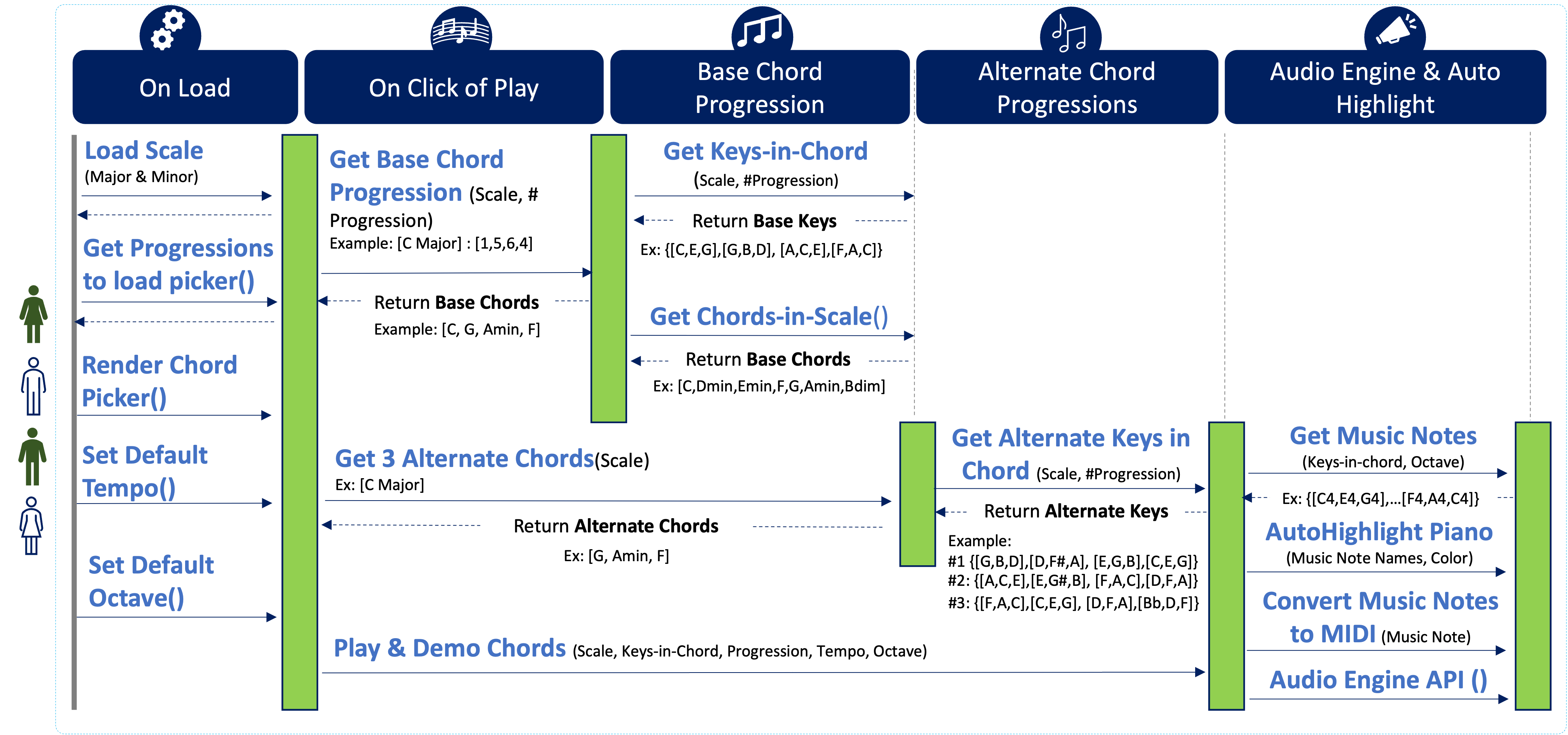}}
\caption{Novel iOS Mobile App Sequence Diagram. This sequence diagram is a detailed depiction of the functions and call hierarchy that programs the Mobile App and User Interface.}
\label{fig}
\end{figure*}

\subsection{Sequence Diagram}
A detailed description of the processes in Fig. 5 is provided below

\textbf{On Load.} On load of the Mobile Application, the 21 Major and 21 Minor scales are loaded in to User Interface PickerViews. PickerViews in the User Interface of SwiftUI are scrollable elements on the screen. These 42 scales in total are hard coded into the system and aren't read from the Java dataset. The 'GetProgressionsToLoadPicker()' function reads through the Java-produced Dataset and accesses the chord progressions. The 'RenderChordPicker()' function uses those accessed progressions by displaying them onto the iOS User Interface. The 'SetDefaultTempo()' selects 120 Beats per Minute as the default tempo out of the Tempo range established prior. The 'SetDefaultOctave()' selects C4 as the default octave out of the Octave range. 

\textbf{On Click of Play Button.} On click of the Play Button, the 'GetBaseChordProgression()' function takes the Scale and Numerical Progression selected by the user to then return the Base Chord progression (User-selected chord progression). The 'Get3AlternateChords()' function takes the Scale selected by the user as an input to make a call for the alternate variations based on the logic described prior. The final call made is from the 'Play\&DemoChords()' function which takes Scale, notes in each chord in the progression, scale progression, tempo, and octave. 

\textbf{Base Chord Progression.} After the call is made from the 'GetBaseChordProgression()' function, the 'GetKeysInChord()' function takes scale and numerical progression as inputs to read the dataset and identify the exact set of notes per chord in the progression. This function will read the dataset and identify the exact combination of scale and numerical progression passed by the user to then output keys. For example, this would output \{[C,E,G],[G,B,D], [A,C,E], [F,A,C]\} for the scale of C major and the numerical progression of 1,5,6,4. Additionally, the 'GetChordsInScale()' function is invoked, returning the 'allowed chords in a scale' which refers to all possible chords in a scale. This is then displayed in the bottom panel as shown in Fig. 4. As was the case with the keys per chord, this process also reads the dataset to ensure that the correct scale is being displayed on the User Interface. 

\textbf{Alternate Chord Progressions. } After the call from the 'Get3AlternateChords()' function on click of the Play Button, the 'GetAlternateKeysInChord()' function is invoked. This function takes Scale and Numerical Progression as inputs and goes to the dataset to identify the appropriate variations based on the user's selection of scale and progression. As such, this segment returns the alternate keys per chord for each of the three alternate variations. 

\textbf{Audio Engine and Auto-Highlight. }  This set of functions is primarily focussed on production sound and conversion from note to MIDI. The 'GetMusicNotes()' function takes the keys-in-chord for the main progression from the user and alternate variations. This function essentially combines each note from the array (i.e. 'C' from \{[C,E,G],[G,B,D], [A,C,E], [F,A,C]\}) with the octave number selected by the user from the stepper between 1-9. The output for this function will be \{[C4,E4,G4],[G4,B4,D4], [A4,C4,E4], [F4,A4,C4]\}. The next function is the 'AutoHighlightPiano()' function. A contribution of this work is a user-friendly interface. In order for this mobile app to truly be 'user-friendly', ensuring functionality for the user to interact with the keyboard while it is playing the chord progressions was important. The design of this interactive piano involves multiple parts. Specifically, the music note names are encoded to represent keys on the keyboard in a swift file. Then, the output from 'GetMusicNotes()' will trigger the color change of those keys indicated in the array such that the interactive keyboard is updated in realtime. The next function 'ConvertMusicNotesToMIDI()' takes each music note in the music notes array and converts it to MIDI based on established MIDI Conversion algorithm. This process converts the labels 'C4' to '60'. The MIDI conversion is important because in the next segment, 'AVAudioEngine()' - Apple's audio API- MIDI notes are passed as inputs along with tempo and octave (selected by the user) and trigger the sound system to play the corresponding pitch. The labels on the piano were timed with the AudioEngine API such that when a key changed color on the keyboard, the AudioEngine would process the corresponding MIDI note. The three pitches (that compose the entire chord) were overlayed through an overlay function for the API. The keyboard was not allowed to move onto the next chord in the progression until the sound finished processing from the previous AVAudioEngine Call. Currently with the 3,297 four-chord progression possibilities, there is marginal latency. This meets the initial design criteria to ensure User Experience, Robustness, and Scale. 
\section{Conclusion}
This two part solution approach was successful in its approach, being the first two-part solution and enumeration algorithm to enumerate every possible four-chord and eight-chord progression within the confines of Music Theory grammar. Additional innovations were made with the addition of an alternate variation for each chord progression selected by the user based on music theory established in \cite{b26}. This customizable and thorough tool is interactive, provides multiple fronts for customizability, and performs with robustness. With the 3,000+ Four-chord progressions and 400,000+ eight chord progressions generated, musicians will now be able to unleash a level of creativity with unique progressions that hasn't been explored prior. Furthermore, the availability of such numerous data regarding chord progressions hasn't been produced before and can drastically help with training machine learning models with chord progressions. The discovery of the chord progressions proposed in this study warrants a future investigation into more chord progression to mood analysis. 

It should be noted however, that although the Java Algorithm successfully enumerated four-chord and eight-chord progressions, the iOS mobile Utility currently only displays the four-chord progressions due to User Interface Constraints. As per the current design, there is a 4x4 matrix that displays chord progressions such that the first row is the main chord progression, while the remaining three rows are the alternate variation progressions. With an eight-chord progression UI, drastic changes to the 4x4 matrix would be needed. 

In the short term, some possibilities for this project could be adding an option for users to save and share MIDI chord progressions to embed in their digital audio workstations. Adding loops for playback could also be of use for future ideas. Additionally, while this chord progression algorithm provides alternate variations to users uniquely, future studies could be employed to study how to generate more alternate variations to allow for further customization for users. Currently, the iOS design displays chord progressions in their root format (ascending scale degrees 1, 3, and 5; or ascending scale degrees 3, 5, and 7). Often times in composition, however, having dynamic ability to play inverted chords (i.e. 3, 5, and 1) allows for layered tonal texture that is appealing. Integrating this functionality in the application would only require a script that identifies when to shift one of the three notes (i.e. C4, E4, and G4) from the fourth octave to the fifth octave. 

In the long term, the data generated from this project can be used for Artificial Intelligence applications. With the new trends in Artificial Intelligence in music to predict moods or human-related events from music, this data can power further exploration of novel progressions with moods. Additionally, in classification machine learning models, these unexplored chord progressions can either be classified on the emotion they evoke, genre, style, or even derive new classifications altogether. In terms of the app, adding the functionality to rearrange chords in a progression, add notes/text editing for lyric generation, or recording melodic ideas could all help with transforming the app from a medium for composition to a Digital Audio Workstation.

\vspace{12pt}

\end{document}